\documentclass{iopart}

 \usepackage{graphicx}

\usepackage{amssymb}

\begin{document}




\title{First excitations in two- and 
three-dimensional random-field Ising systems}


\author{M Zumsande$^1$, M J Alava$^2$ and A K Hartmann$^3$}

\address{$^1$ Institut f\"ur Theoretische Physik, Universit\"at
G\"ottingen, Friedrich-Hund-Platz 1, 37077 G\"ottingen, Germany}
\address{$^2$ Laboratory of Physics, Helsinki University of
Technology, 02015 HUT, Espoo, Finland} \address{$^3$ Institute of
Physics, University of Oldenburg,  26111 Oldenburg, Germany}
\eads{\mailto{zumsande@theorie.physik.uni-goettingen.de},
\mailto{mja@fyslab.hut.fi}, \mailto{a.hartmann@uni-oldenburg.de}}

\begin{abstract}
We present results on  the first excited states for the random-field
Ising model. These are based on an exact algorithm, with which we
study the excitation energies and the excitation sizes for two- and
three-dimensional random-field Ising systems with a Gaussian
distribution of the random fields.  Our algorithm is based on an
approach of Frontera and Vives which, in some cases, does not yield
the true first excited states. Using the corrected algorithm,
we find that the order-disorder phase
transition for three dimensions is visible via crossings of the
excitations-energy curves for different system sizes, while in
two-dimensions these crossings converge to zero disorder.  Furthermore, 
we obtain
in three dimensions a fractal dimension of the excitations cluster of
$d_s=2.42(2)$. We also
provide analytical droplet arguments to understand the behavior of the
excitation energies for small and large disorder as well as
 close to the critical  point.
\end{abstract}


\section{Introduction}
\label{s:intro}
Many disordered systems are characterized by  an
extremely rough energy landscape \cite{Young98,Fischer91,Mezard87}.
This leads to the existence of
configurations that are quite different from ground-state configurations
but have an energy very close to the ground-state energy. 
The behavior of such systems differs considerably from the behavior
of ordered systems and has attracted the attention of many physicists 
during the last decades.
The most prominent example of a magnetic system with quenched disorder
exhibiting many valleys in the energy landscape is the spin-glass model
\cite{Binder86}.
 
Another basic model from statistical physics that exhibits quenched disorder is
the random-field Ising model (RFIM). In the RFIM, the sensitivity to
the actual configuration of the disorder and the roughness of the
energy landscape are not as strong as in spin glasses but nevertheless
present and important to understand the behavior \cite{Alava98}. 

A RFIM sample consists of $N=L^d$ Ising spins $s_i$ on a
$d$-dimensional regular hypercubic lattice with lateral length $L$ in
each dimension. Neighbouring spins interact ferromagnetically with
strength $J$. In addition, on each site  a random magnetic field $h_i$
acts on spin $s_i$.  The $h_i$ are quenched random variables chosen
according to some probability distribution. One common choice is a
Gaussian normal distribution of zero mean and standard deviation
$h$. With a Gaussian disorder, the ground state and the FES are
nondegenerate.

The Hamiltonian of the RFIM is
\begin{equation}
\label{eq:rfim}
H=-J \sum_{\langle ij \rangle} s_i s_j - \sum_{i=1}^N h_i s_i.
\end{equation}

The current state of knowledge on the Gaussian RFIM can be sketched as
follows: No ordered phase can persist in the infinite system for
$d<3$, the lower critical dimension is $d_l=2$ \cite{Bricmont88}. For $d=3$, there is a continous phase transition at a critical field $h_c$ 
between an ordered ferromagnetic phase and a disordered paramagnetic
 phase \cite{Middleton01}.
It is possible to analyse the phase transition at zero temperature since the
random fields are a relevant perturbation as shown by Fishman and
Aharony \cite{Fishman79}.

Compared to other disordered models such as spin glasses, the RFIM has
the practical advantage that in any dimension the  problem of
finding the ground state can be solved with fast graph-theory based
algorithms. This means that the running time increases only like a
low-order polynomial in $N$.  This has led to numerous studies of the
ground state landscape of the RFIM,
e.g. \cite{Middleton01,Hartmann99,Hartmann02}. The system sizes
that can be analyzed this way are considerably larger than those that
can be equilibrated via Monte-Carlo techniques at small temperatures
\cite{Newman96,Rieger95}.

Using these approaches one is not restricted to study only ground states.
Even better, it is also possible to study the properties of 
the RFIM at very low
but nonzero temperatures with suitably modified
ground-state techniques. To do this, the
ground state must be perturbed in some way to gain low energy
states. Various attempts have been made in this direction, among them
Refs.\ \cite{Alava98,Middleton01}.

In this study, we investigate 
 first excited states (FESs) in the RFIM and their
differences to ground states. In particular we present 
in Sec.\ \ref{s:algo} an algorithm
to obtain exact FESs, in contrast to a previous attempt \cite{Frontera01},
where some FESs were missed. In Sec.\  \ref{s:results},  we show first 
that indeed
a considerable fraction of true FESs was missed in Ref.\ \cite{Frontera01}.
Next we apply our algorithm to study FESs both in the
ordered and disordered phase. In particular, we investigate the
scaling behavior of the excitation energy and the fractal properties
of the FESs. These findings are finally contrasted with 
arguments based on extreme-value considerations and on the
droplet theory for the excitations in a hierarchical landscape.
Finally, we finish with conclusions.

\section{Algorithm}
 \label{s:algo}
A fast ground-state algorithm for the RFIM was first suggested in Ref.\
\cite{Picard75} and is explained in detail in Ref.\ \cite{Hartmann01}. The main
idea is to transform the RFIM to a {\em network} ${\mathcal N}=(G,c,s,t)$ consisting of
a directed, weighted graph $G=(V,E)$. $V$ is the set of nodes and
$E\subset V\times V$ being the set of edges. $c_{i,j}\ge 0$
 are edge labels denoting capacities.  There are two distinguished 
nodes $s$ and $t$ called the source and the sink. All other nodes
$i\neq s, i\neq t$ are called {\em inner} nodes. In
${\mathcal N}$, a $(s,t)$-cut $X=(S,\overline{S})$ is defined as a partition of
$V$ into two disjoint sets $S$ and $\overline{S}$
with $s\in S$ and $t \in \overline{S}$. 
We call  $S$ and $\overline{S}$ the two {\em sides} of the cut.
To each $X$, a capacity
\begin{equation}
C(X)=\sum_{i\in S,j\in \overline{S}} c_{ij}
\label{eq:cutcap}
\end{equation}
is assigned. Note, this means that only edges with the ``right''
direction $i \to j$ contribute to $C(X)$. We will also use an equivalent 
description of a cut by a vector $X_i\in \{0,1\}$ 
($i \in V$) with
$X_i=0$ if $i\in S$ for $X$ and $X_i=1$ else.

The problem of finding the ground state of the RFIM can be reduced to
the problem of finding the cut $X^{min}=X^{min}({\mathcal N})$ 
with the smallest capacity
among all possible cuts in such a network. This problem can be solved
by determining the maximum flow of the network, as proven by Ford and
Fulkerson \cite{Ford56}. As already mentioned, the ground state is nondegenerate, hence
 $X^{min}$ is also unique for the model with Gaussian distribution of the random
fields.

We will now describe the algorithm of finding the first excited state
as developed by Frontera and Vives (FV), adopting most of their
terminology. By means of an example, we show why there are cases in
which this algorithm fails to find the FES. The necessary adjustments
are subsequently explained.

Let ${\mathcal N}$ be a network corresponding to a RFIM sample.  Its minimum cut
$X^{min}$ can be calculated with a maximum-flow algorithm.  
We say an edge $w=(i,j)$ is {\em contained} in $X$, if
$i\in S$ and $j\in\overline{S}$.
We denote the set of edges that are contained in $X$
as $E_c(X)=\{(i,j)\in E|i\in S, j\in\overline{S}\}$
and write $w \in E_c(X)$.  The set of nodes that are on
\emph{either side} of any edge $w \in E_c(X)$ is called $V_c(X)$,
i.e.\ $V_c(X)=\{i\in V|\exists\, j\in V: (i,j)\in E_c(X)\,\mbox{or}\, 
(j,i)\in E_c(X)\}$.

The set of all possible cuts (corresponding to all possible
spin configurations) of ${\mathcal N}$ is $S^{\mathcal N}$. Any edge $w$ divides $S^{\mathcal N}$
into two disjoint subsets, the set $S^{\mathcal N}_w$ of cuts that contain $w$
and the set $\bar{S}^{\mathcal N}_{w}$ that do not.

Next, a second network ${\mathcal N}^\delta_{\bar{w}}=(G,c^\delta_{\bar{w}},s,t)$ 
is defined that includes a
perturbation of strength $\delta$. The difference between
${\mathcal N}^\delta_{\bar{w}}$ and ${\mathcal N}$ is that the capacity of one edge $\bar{w}
\in E_c(X^{min})$ of the minimum cut is increased by an amount $\delta
\gg 0$, so that
\begin{eqnarray}
c^\delta_{\bar{w}}(\bar{w})&=&c(\bar{w}) + \delta \nonumber \\
c^\delta_{\bar{w}}(w)&=&c(w) \hspace{2cm} \forall w \neq \bar{w}.
\end{eqnarray}
Each cut $X$ of ${\mathcal N}$ is also well defined in ${\mathcal N}^\delta_{\bar{w}}$, but
its capacity has increased if $\bar{w}$ is contained in $X$. For the
remaining cuts of ${S}^{\mathcal N}_{\bar{w}}$ that do not contain $\bar{w}$, the
capacity does not change, i.e.
\begin{equation}
C^\delta_{\bar{w}}(X)-C(X)=\left\{ \begin{array}{rl}
0 &\mbox{if $\bar{w} \notin X$} \\
\delta &\mbox{if $\bar{w} \in X$} 
\end{array}
\right.
\end{equation}
Since we chose $\bar{w} \in E_c(X^{min})$, $X^{min}$ is the cut with
the minimum capacity in $S^{\mathcal N}_{\bar{w}}$ for both ${\mathcal N}$ and
${\mathcal N}^\delta_{\bar{w}}$ (the capacity of all cuts of $S^{\mathcal N}_{\bar{w}}$ is
increased by $\delta$ in ${\mathcal N}^\delta_{\bar{w}}$).  The minimum cut of
the complementary set $\bar{S}^{\mathcal N}_{\bar{w}}$ is called
$\bar{X}^{min}$. For $\bar{X}^{min}$ and all other cuts of this set,
the capacity is the same in both networks.\\ It follows that depending
on $\delta$, there are two possibilities for the minimum cut of
${\mathcal N}^\delta_{\bar{w}}$:
\begin{equation}
X^{min}({\mathcal N}^\delta_{\bar{w}})=\left\{ \begin{array}{rl}
X^{min} &\mbox{if } C(X^{min})+ \delta < C(\bar{X}^{min}) \\ 
\bar{X}^{min} &\mbox{if } C(X^{min})+\delta > C(\bar{X}^{min})
\end{array}
\right.
\end{equation}
If $\delta$ is chosen large enough (e.g. $\delta=f^{GS}_{max}$, the
maximum flow in the ground state), the minimum cut of ${\mathcal N}^\delta_{\bar{w}}$ is
always $\bar{X}^{min}$.

Let $X^{2nd} \in S^{\mathcal N}$ be the cut with the second smallest capacity in
${\mathcal N}$, corresponding to the first excited state. This cut must differ
from the minimum cut by at least one edge. Frontera and Vives claim
that because of this, ``there exists (at least) one edge $w$ of
$X^{min}$ so that $X^{2nd}$ is in $\bar{S}^{\mathcal N}_w$''. We will show below
that this assumption is not always true.  

To understand the algorithm of Frontera and Vives, we assume for a moment
that the assumption is true.
Then, in order to find the
capacity of the second minimum cut, one would have to choose a suitably large
$\delta$. After that, $C(\bar{X}^{min})$ is determined for all
networks ${\mathcal N}^\delta_{\bar{w}}$ depending on the edges $\bar{w}$
contained in the minimum cut $X^{min}$.  From all the resulting cuts,
the one with the minimum capacity in the original network ${\mathcal N}$ is
accepted and called $C^*$.
\begin{equation}
C^*=\mbox{min}_{\bar{w} \in E_c(X^{min})}C\left(\bar{X}^{min}\right).
\end{equation}

However, the FES has not necessarily been found yet, $C^* \neq
C^{2nd}$ since the algorithm still has a problem: It assumes that the
minimum cut necessarily includes an edge that the second minimum cut
does not include. This is wrong, since there is also the possibility
that all edges of the minimum cut are still contained in the second
minimum cut which differs from the old one in that it
\emph{additionally} contains another edge.

\begin{figure}[h]
  \centerline{\includegraphics[width=70mm]{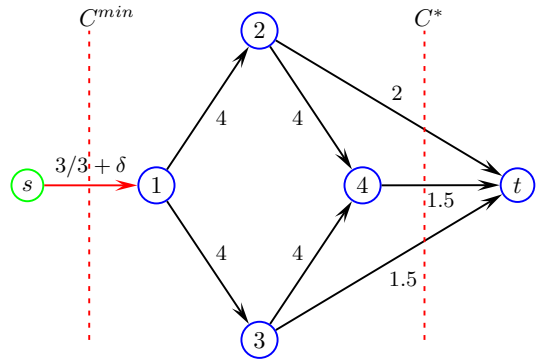}}
  \caption{Minimimum cut $X^{min}$ and $X^*$ for an example network}
  \label{g:exrfimviv}
\end{figure}

\begin{figure}[h]
  \centerline{\includegraphics[width=70mm]{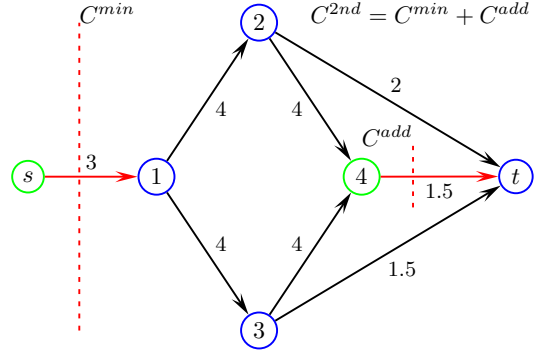}}
  \caption{Second minimum cut $X^{2nd}$ for the example RFIM.}
  \label{g:exrfimfes}
\end{figure}

This can be illustrated by looking at Fig.\ \ref{g:exrfimviv}, an example
network where the GS cut is drawn in. When the capacity of the edge connecting $s$ with 
node $1$ is enlarged by a sufficiently large amount $\delta$, the flow
through the new network is limited by the edges to the
sink. Therefore, in the resulting cut $X^*$ that is shown in the figure, 
all inner nodes have changed their side of the cut, and the capacity of $X^*$ is 
$C^*=2+1.5+1.5=5$.  $X^*$ is not identical with the true
second minimum cut $C^{min}$ that is depicted in
Fig.\ \ref{g:exrfimfes} and has a capacity $C^{min}=3+1.5=4.5$, so that
$C^{min}<C^*$.

We change the algorithm by leaving the original algorithm untouched
(obtaining an intermediate cut $C^*$ after the first step) and
introducing a second step.  In this step, we go through all nodes $k\in V$ 
except those belonging to $V_c(X^{min})$.  For each of these nodes $k$, 
we consecutively check if it is favorable to assign $k$
to the other side of the cut (in the vector notation of a cut, set
 $X_{k}^{new}=0$ if $X_{k}^{min}=1$ and vice versa). 
This would result in a new cut $X^{new}$ for
which $E_c(X^{min}) \subset E_c(X^{new})$. As a result, additional
edges contribute to the capacity of $X^{new}$. We denote their
contribution with $C^{add}$ so that $C^{new}=C^{min}+C^{add}$.
Note that performing this test for a node 
belonging to $V_c(X^{min})$ would result also in removing some edges
from the cut, hence this is already taken care of by the FV
algorithm.

If there are  nodes $k$ where $C^{new} < C^*$, the FES is given by the
minimum cut among the nodes $k$, if no such node exists, the FES is $X^*$.

Note that there can never be a negative contribution to $C^{add}$
since with $c_{ij} \geq 0 \quad \forall i,j$ this would mean to remove
an edge that was included in $X^{min}$ in the GS (while adding other edges at the same time).
But in this case the removed edge would have been already considered in the first step.

For the extension of the FV algorithm, it is sufficient to look at single
spin flips, i.e.\ at moving exactly one node from one side of the cut 
to the other. We explain this for a node $k \in V$ with $X^{min}_k=0$, the
other case $X^{min}_k=1$ can be treated analogously.  The change 
$X^{min}_k=0\,\to\, X^{new}_k=1$ 
for $k \notin V_c(X^{min})$ affects the
capacity of the new cut in the following way:\\
 There are possibly nodes
$X^{min}_{i}=0$ and $c_{ki}>0$. In this case 
an amount of $c_{ki}$ is added to $C^{add}$ as soon as $k$ is set to $1$.
Since $k\notin V_c(X^{min})$ there are no predecessor nodes 
$i$ with $X_i^{\min}$ =1,
hence incoming edges into $k$ do not have to be considered for
the calculation of $C^{add}$, due to (\ref{eq:cutcap}).

It might seem plausible that there exist situations where, instead of
adding the capacities $c_{ki}$ to $C^{add}$, also node $i$
(and possibly more nodes) should be moved to the other side of the
cut. In fact, some of these situations would lead to a smaller
$C^{add}$. However, in such a situation it would lead to an even
smaller $C^{add}$ (remember that we are only interested in the second smallest cut)
to invert \emph{only} node $i$ instead of inverting
both node $k$ and node $i$. This is due to the fact that with our
method, there can never be a negative contribution to $C^{add}$.

The pseudo code for the full algorithm for calculating the FES 
is shown in Fig.\ \ref{fig:alg}.

\begin{figure}
\begin{tabbing}
\textbf{algorithm} First Excited State\\
\quad Transform the RFIM to a network ${\mathcal N}$\\
\quad Calculate $X^{min}$ of ${\mathcal N}$ that corresponds to the ground state\\
\quad $C^{2nd}:=C^{min}+\delta$ with $\delta=f^{GS}_{max}$\\ 
\quad \textbf{for} each edge $w \in X^{min}$
\quad  $\{$ Frontera-Vives algorithm $\}$ \\
\qquad $c(w):= c(w)+\delta$ \\
\qquad $X_{w}:=$ minimum cut of ${\mathcal N}^\delta_w$\\
\qquad  \textbf{if} $C^\delta_w(X_w)<C^{2nd}$ \textbf{then}  \\
\qquad \quad  $C^{2nd}:=C^\delta_w(X_w)$ \\
\qquad \quad $X^{2nd}:=X_w$\\
\qquad \textbf{end if}\\
\qquad $c(w):= c(w)-\delta$ \\
\quad \textbf{end for}\\
\quad \textbf{for} all nodes $k \notin V_c(X^{min})$  
\quad $\{$ extension to obtain true FES $\}$\\
\qquad $C^{add}:=0$\\
\qquad \textbf{if} $X^{min}_k=1$ \textbf{then}\\
\qquad \quad \textbf{for} all nodes $i$ with $c_{ki}>0$\\
\qquad \qquad $C^{add}:=C^{add}+c_{ki}$\\
\qquad \quad \textbf{end for}\\
\qquad \textbf{else if} $X^{min}_k=0$ \textbf{then}\\
\qquad \quad \textbf{for} all nodes $i$ with $c_{ik}>0$\\
\qquad \qquad $C^{add}:=C^{add}+c_{ik}$\\
\qquad \quad \textbf{end for}\\
\qquad \textbf{end if}\\
\qquad \textbf{if} $C^{min}+C^{add}<C^{2nd}$\\
\qquad \quad $C^{2nd}:=C^{min}+C^{add}$\\
\qquad \quad $X^{2nd}:=X^{min}$ with node $k$ inverted\\
\qquad \textbf{end if}\\
\quad \textbf{end for}\\
\quad \textbf{return} $C^{2nd}$ and $X^{2nd}$ \\
\textbf{end}
\end{tabbing}
\caption{Pseudocode of the full algorithm to obtain the exact FES. 
\label{fig:alg}} 
\end{figure}

Finally, we discuss the running time of the algorithm.
The running time of the  FV algorithm
 scales like \cite{Frontera01} 
$t \sim N^{\alpha+1}$
when the running time of the minimum cut has scales as $t \sim
N^\alpha$.
The additional part introduced here
 is linear in $N$ and therefore does not increase
the full running time $t$ of the algorithm in a noticeable way. 

\section{Results}
\label{s:results}
We calculated first excited states of $10^4$ samples at different
random field strengths $h$ and different system sizes up to $L=16$ in
$d=3$ (resp. $L=32$ in $d=2$). For the largest three-dimensional
system size $L=20$, $2 \cdot 10^3$ samples were calculated at each
$h$.

A comparison of the states generated with the Frontera-Vives algorithm
with those generated using the complete algorithm that is described above,
 shows that the differences are
noticeable though not drastic. For small systems in $d=2$ with $N=4^3$ spins at
$h=2.3$, the FV algorithm does not find the
true FES in $15\%$ of the cases. With growing
system size, this percentage decreases. In two-dimensional $32\times 32$
samples (systems of this size were mainly investigated in \cite{Frontera01}),
the percentage is $5\%$ and in
three-dimensional systems of $L=16$, $1\%$ of the excitations
generated by the FV algorithm are
incorrect - both in the ferromagnetic and paramagnetic region. In the
transition region between the two phases, the error rate is considerably smaller
since a large number of first excited states consist of more than one
spin and are therefore spotted correctly by the FV algorithm.

\begin{figure}[h]
\centerline{\includegraphics[width=100mm]{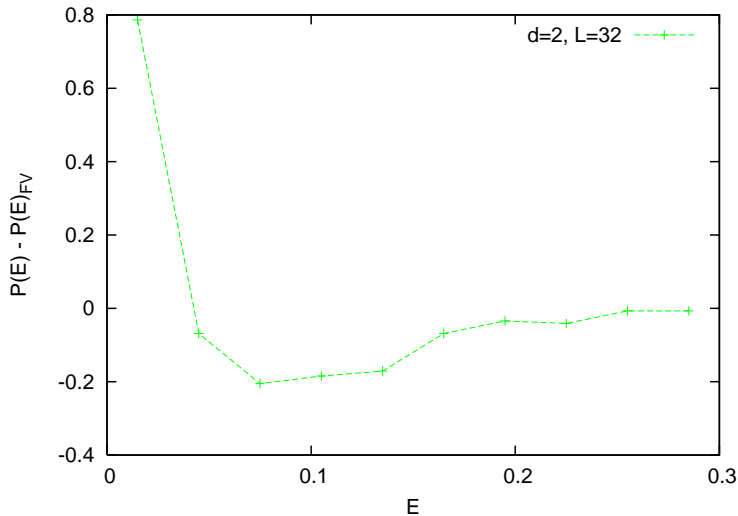}}
\caption{Difference $P(E)-P(E)_{FV}$  between excitation energy distributions for the complete algorithm and the FV algorithm. Lines are guides to the eyes
only.}
\label{g:fescompare}
\end{figure}

Since the FV algorithm creates states with a higher excitation energy, the difference is also visible in the distributions of the excitation energy. The difference $P(E)-P(E)_{FV}$ is positive for small $E$ and negative for larger $E$, as shown in Fig.\ \ref{g:fescompare}.

In our analysis of the properties of the FESs, the main quantities of
interest are the energy difference $E$ to the GS and the size $V$ of the
excitation cluster, given by the number of spins that change their
orientation in the FES.
By plotting the disorder-averaged excitation energy $\overline{E}$ against $h$ for
various $L$, we see that for large $h$, the $E(h)$ becomes independent
of $h$, i.e.\ $\overline{E}(h)\to \mbox{const}(L)$. The value of the plateau scales
like $\overline{E}(h\to \infty) \sim
L^{-d}$.

\begin{figure}[h]
\centerline{\includegraphics[width=100mm]{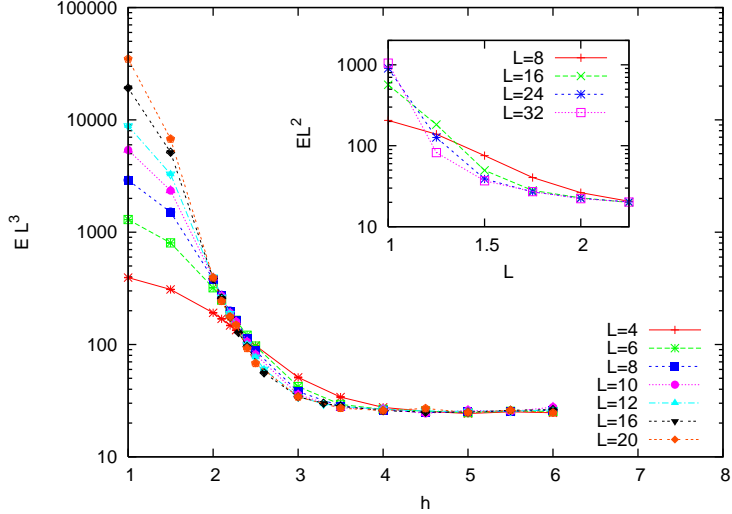}}
\caption{Excitation energy versus random field strength. Inset: $\overline{E} \cdot L^2$ versus $h$ for $d=2$. Main part: $\overline{E}
\cdot L^3$ versus $h$ for $d=3$.  Lines are guides to the eyes
only.}
\label{g:fespap23d}
\end{figure}

For that reason, we rather plot $\overline{E}L^d$ instead of $\overline{E}$ vs. $h$. The
resulting curve for $d=3$ is shown in the main part of
Fig.\ \ref{g:fespap23d}, the corresponding curve for $d=2$ is plotted
in the inset.

For $h \gg h_c$, the curves of all system sizes coincide as expected.
Slightly above the phase transition, the rescaled energy grows
with decreasing $h$.
 For larger systems, the slope is larger. An
interesting point is that in $d=3$ the different curves intersect
 in a narrow region $2.1 \leq h \leq 2.3$ that is close
to $h_c$. This behavior can only be seen in three dimensions. In $d=2$
there is no
single intersection point and the intersections approache 
$h=0$ with increasing $L$, reflecting the absence of a phase
transition in two-dimensions \cite{Bricmont88}. Hence, it seems to
be possible to locate the phase transition from the scaling behavior of the
excitation energy. We are not aware of previous similar findings.
In the next section, we will show by simple droplet arguments,
that indeed the energy close to $h_c$ should display a $L^{-d}$
behavior. Nevertheless, we are not able to give a full
explanation of this behavior around $h_c$, which seems to originate
from corrections to scaling.

\begin{figure}[h]
\centerline{\includegraphics[width=100mm]{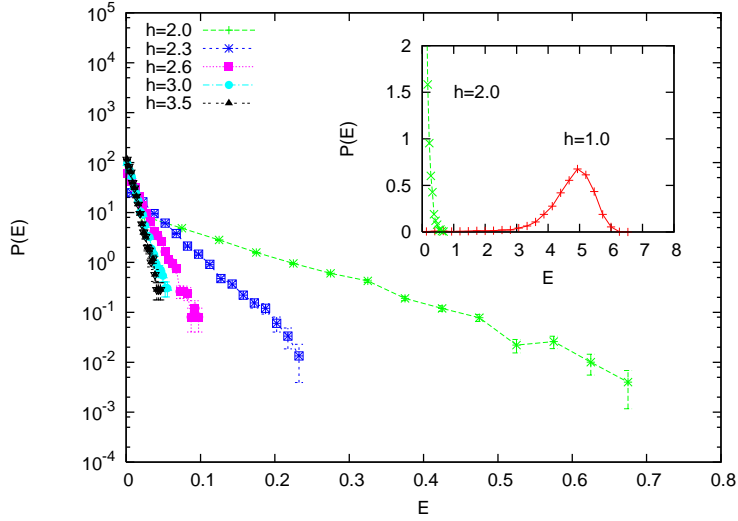}}
\caption{Probability distributions of the $3d$ excitation energy at
fixed system size $L=16$ for various $h$. In the inset, the range of
the x-axis is $[0:8]$, in the main part, the region of small $E$ in
the range $[0:0.8]$ is shown.  Lines are guides to the eyes
only.}
\label{g:multipe}
\end{figure}

The energy distributions $P(E)$ are shown in Fig.\ \ref{g:multipe} for
fixed $L=16$. Close to and beyond $h_c$ the distribution seems to
be exponential, as we also find using the simple droplet arguments
in the next section.
In the inset, the extremely ferromagnetic case $h=1.0$
is depicted. $P(E)$ is peaked at a large $E$ in completely
ferromagnetic samples, corresponding to flipping
spins against the orientations of all neighbors.
For a detailed discussion, see again below. Small excitations cannot occur,
there is an
\emph{energy gap} between the GS and the FES.

\begin{figure}[h]
\centerline{\includegraphics[width=100mm]{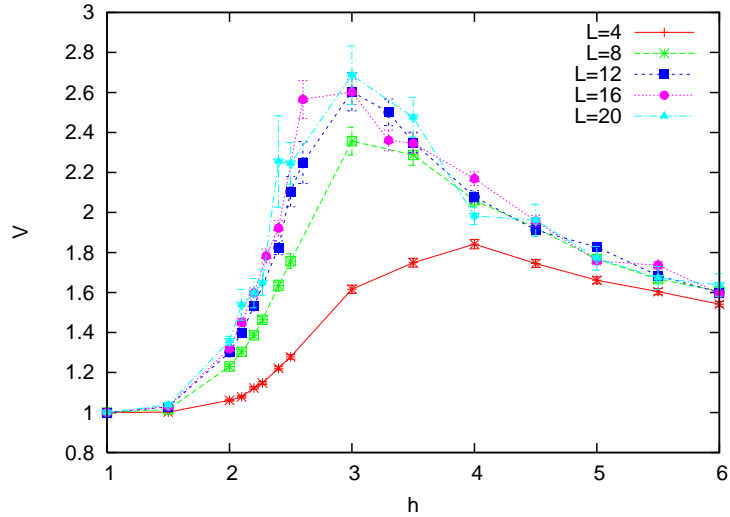}}
\caption{Average volume of the FES dependent on $h$ for various $L$. Lines 
are guides the eyes only.}
\label{g:fesvmean}
\end{figure}

The average size $V$ (i.e.\ the number of flipped spins) of the FES is
shown in Fig.\ \ref{g:fesvmean}. Global flips were not taken into
account for this plot.
In the ferromagnetic phase, $V$ is equal to one. For $h\to \infty$, it
also decreases slowly towards $V=1$. Thus, in both cases, 
the FES behaves as expected above.
In between, there is a maximum
with a size that seems to grow with $L$. The value of $h$ where the maximum
is located depends on $L$. It shifts towards $h_c\approx 2.27$ with
growing system size. This is due to the fact that finite systems are
most ``critical'' at a field $h_c(L)>h_c$ that approaches $h_c$ when
$L\to \infty$. Since for first excited states, we are restricted to
rather small system sizes and the statistical 
errors are still significant, it remains unclear
if the maximum really moves to $h_c$ for larger systems.

\begin{figure}[h]
\centerline{\includegraphics[width=100mm]{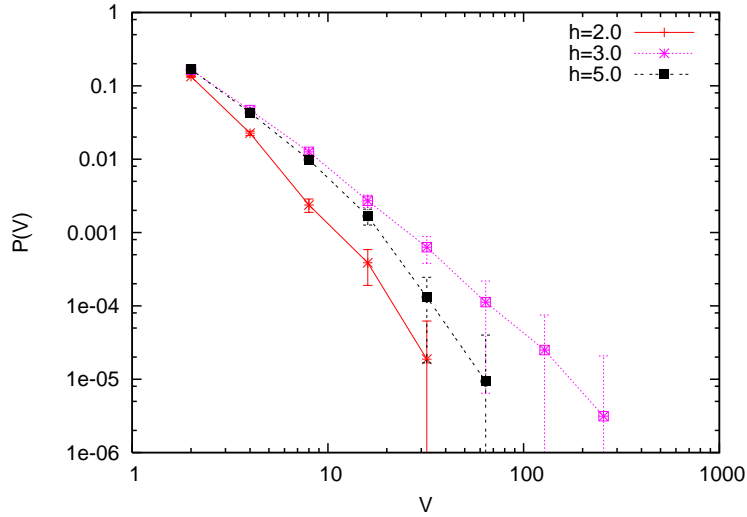}}
\caption{Probability distribution $P(V)$ at fixed $L=16$ and
random-field strengths $h=2.0$, $h=3.0$ and $h=5.0$. Lines are guides 
to the eyes only.}
\label{g:volhist}
\end{figure}

The probability distribution of the excitation volumes is shown in the
double logarithmic plot of Fig.\ \ref{g:volhist}. Both in very
paramagnetic ($h=5.0$) and ferromagnetic ($h=1.0$) systems, $P(V)$
declines in a fast way. At $h=3.0$ near the peak of
Fig.\ \ref{g:fesvmean}, the decay is not so fast and resembles at least
for small $V$ a power law. However, to verify that $P(V)$ approaches a
power law at $h_c$ for large systems, much larger systems would be
necessary.

\begin{figure}[h]
\centerline{\includegraphics[width=100mm]{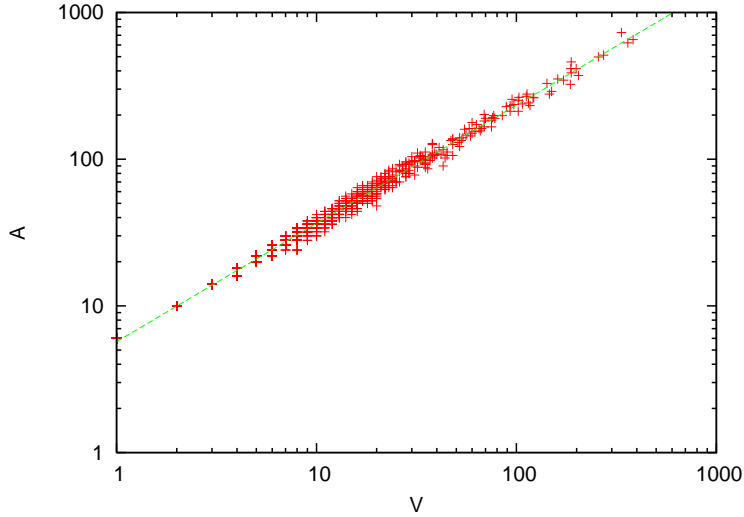}}
\caption{Fractal dimension $d_s$ of first excitation droplet ($L=12$)
at $h=3.0$. Each droplet is characterized by a data point. A 
least-square fit yields $d_s=2.43(2)$.}
\label{g:fesvofa}
\end{figure}

We also analyzed the shape of the excited clusters. The surface $A$ of
a cluster is given by the number of bonds that connect a spin of the
cluster with another spin which is not part of the cluster. By
plotting $A$ against the volume $V$ of the clusters, we gain
information on the fractal dimension $d_s$ of the cluster. Since we
found that the clusters are compact ($V \sim R^3$, not shown) 
it follows that $A\sim V^{d_s/d}$.  From Fig.\ \ref{g:fesvofa} ($h=3.0$, $L=12$) it
follows that $d_s=0.808$, leading to $d_s=2.43(2)$ which can be compared
with an earlier result $d_s=2.30(4)$ of Middleton and Fisher 
\cite{Middleton01} obtained for domain-walls. 

\section{Extreme-value and droplet arguments for the minimal excitations}

Now, we try to understand the behavior of the distribution of the
droplet energy and the average
via analytical arguments for the extreme cases $h\ll h_c$
and $h\gg h_c$ as well as in the region where the droplet theory
holds, hence close to $h_c$. 

The relation $\overline{E} \sim N^{-1}$ for large $h$ can be explained by a simple
extreme-value 
argument, which e.g.\ has also been used to explain the 
low-temperature behavior of spin glasses
\cite{Bray84}.
 If $h \to \infty$, the system behaves paramagnetically. Hence, 
FESs will be essentially single-spin flip excitations which is indeed true 
for the simulation data, as we have seen above.
In this case, the spin with
the smallest total local field will flip for the FES.
Due to the paramagnetic state,
there will be many spins with the same number of ``up'' neighbors and
``down'' neighbors. Hence, the total local field will be the same as
the quenched local field $h_i$. The energy of the 
FES will be given by the smallest absolute value $\vert h_i \vert$
among these spins.\footnote{There might be also spins with the number
of ``up'' spins not equal to the number of ``down'' spins, but with the
total local field close to zero. Nevertheless, since the Gaussian is 
peaked at zero, those will much less frequently comprise the FES compared 
to the spins discussed here.}
Thus, the excitation energy is
given by the minimum of $O(N)$ absolute values of
Gaussian random numbers. For simplicity, we assume $O(N)=N$:
\begin{equation}
\vert h_{eff} \vert= \vert E \vert/2= \min \{\vert h_1 \vert,\vert
h_2\vert ,...,\vert h_N\vert \}
\end{equation}
The distribution function $P_N(\vert h_{eff} \vert)$ of the sample
minimum of $N$ numbers from a distribution $P(h)$ with $P(h)=0$ for
$h<0$ is
\begin{equation}
P_N(\vert h_{eff} \vert)=N P(\vert h_{eff} \vert) \left(1- \int_{0}
^{\vert h_{eff} \vert} dh P(h) \right)^{N-1}.
\end{equation}
If $P(h)$ is continuous and has a finite weight $P(0)$ at the origin
(which is true for a Gaussian distribution), one can approximate
$P(h)$ by $P(0)$ for large $N$ and gets
\begin{eqnarray*}
P_N(\vert h_{eff} \vert) & \approx & N P(0) \left(1- P(0) \vert h_{eff}
\vert \right)^{N-1} \\ 
& = & N P(0) \exp\left((N-1)\ln(1-P(0)\vert h_{eff} \vert )\right) \\ 
&  \approx &  \lambda(N) \exp \lbrace - \lambda(N)
\vert h_{eff} \vert\rbrace
\end{eqnarray*}
with $\lambda(N) =P(0) N$, having the mean $
E=\overline{P_N}(\vert h_{eff} \vert)=\frac{1}{\lambda}\sim N^{-1}$

$E \sim N^{-1}$ agrees with the numerical simulation data of the mean
energy and the predicted exponential decay of $P(E)$ occurs indeed in the numerical data and
can be seen in Fig.\ \ref{g:multipe}

The limit of strongly ferromagnetic systems can also be understood. In
the ferromagnetic phase, there are two possibilities for the FES:\\ In
smaller samples, the entire system can flip. This
happens if the excess of the random fields $\sum_i h_i$ is even
smaller than the minimum bond energy that has to be invested for
flipping a single spin. If all spins flip, the bond contribution of
the total energy stays invariant and the random field contribution
$E_h=\sum_{i} h_i \sim L^{d/2}$ changes its sign, so that the
excitation energy is $E=2 E_h$. However, this type of excitation
cannot persist in large systems, since $E_h \sim L^{3/2}$ becomes soon
larger than the energy of a second type of excitations, the local
reversal of a single spin. This energy decreases for larger systems:

If the system is fully magnetized,  all spins are pointing in the same 
direction. 
We discuss the case that all spins are pointing ``down'', the
reverse case is similar.  Hence, we have
$2|h_i|\le 12$ for all spins.
Then,  from the
$\frac{N}{2}$ spins where the local field is positive,  the
spin with the maximum local field $\max\{h_i\}$ will have the smallest
local field, i.e.\ will be chosen to be flipped for the FES.
 Since
six bonds of strength $J$ from $s_i$ to its neighbors are no longer
satisfied, the excitation energy is $\overline{E}=12J-2 \max\{h_i\}$ in three
dimensions. So we seek for the sample maximum in the ferromagnetic
case.

The sample maximum of $N$ values drawn from a distribution $P(h_i)$
has the distribution
\begin{equation}
P_N(h_{max})=N P(h_{max}) \left( \int_{-\infty}^{h_{max}} dh P(h)
\right)^{N-1}.
\end{equation}

We calculated the integrals numerically and determined the expected
value $\overline{P_N}(h_{max})$ for sample sizes up to $N=1000$. It
was found that $\overline{P_N}(h_{max})$ grows in a logarithmic way
with $N$. Therefore we expect that $12-\overline{E}$ will also grow in a
logarithmic way in systems that are large enough that global spin
flips are negligible.

\begin{figure}[h]
\centerline{\includegraphics[width=100mm]{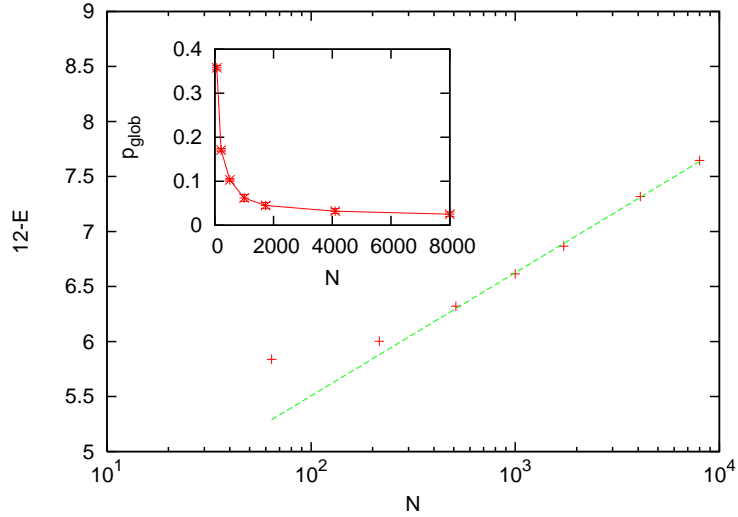}}
\caption{Inset: Probability of global spin flips dependent on the
number of spins $N$. Main part: $12-\overline{E}$ versus $N$. The line 
represents a simple logarithmic function. }
\label{g:fesferro}
\end{figure}

The comparison with the numerical data of the FES deep in the
ferromagnetic phase at $h=1.0$ is shown in Fig.\ \ref{g:fesferro}. In
the inset, the probability of a global flip is plotted versus
$N$. This probability decreases quickly, and for $L=8$, global flips
occur already in less than $5\%$ of the samples. In the main part,
$12-E$ is plotted versus $N$. $12-\overline{E}$ grows in a logarithmic way as
expected, and deviations occur only for the smallest samples where
global spin flips are still frequent.

The droplet argument for the minimal excitation, which should
be valid close to $h_c$, starts with the
assumption that the contributions from various scales are
independent. Further, one takes their (scale-dependent) energy
density to scale as $a/L^\theta$, where $\theta$ is the energy
fluctuation exponent \cite{Fisher88}. The value in 3D is currently
found to be by ground-state simulations to be $\theta =1.49 \pm
0.03$ \cite{Middleton01}. On scale $l = 2^i$, for small energies $E$,
 one will have thus the
cumulative excitation energy distribution $P_l \equiv aE/l^\theta$.

The smallest excitation energy in the whole system is then given by
the mini\-mum over all the $l=1 \dots L = 2^M$ scales. At each scale,
the system splits into $N(l) =(L/l)^D$ independent subvolumes. The
minimum excitation energy from a scale $l$ has the cumulative
distribution
\begin{equation}
P(E,l) = 1-(1-P_l)^{n(l)} \label{lscale}
\end{equation}
where $n(l) = (L/l)^D$.
This comes simply from the argument that the complement of
$P(E,l)$ indicates that all the subvolumes have a smallest 
excitation of higher energy than $E$. Similar ways of
reasoning have been used in other disordered systems \cite{Middleton01b}.

The minimal energy (cumulative) distribution over all the scales can be
constructed similarly. The droplet enery is smaller or equal
to $E$ if the converse is not true: that all the scales would
have a minimum higher than that. Thus we obtain
\begin{equation}
P(E) = 1-\prod_l (1-P(E,l)) = 1 - \prod_l (1-P_l)^{n(l)}.
\end{equation}
The product is now taken over logarithmic scales, so that
$1 \leq l \leq e^{\ln L}$. 

The product can be converted by standard tricks to a
sum  and then to an integral, or computed numerically
from
\begin{equation}
P(E) = 1 - \exp{\sum_{l=1 \dots \ln L / \ln 2}
n(l) \ln (1-P_l)}
\label{eq:PEfinal}
\end{equation}
where one notes that the argument of the exponential has a prefactor
$L^d$ coming from the $n(l)$. 

\begin{figure}
\begin{center}
\includegraphics[width=9cm]{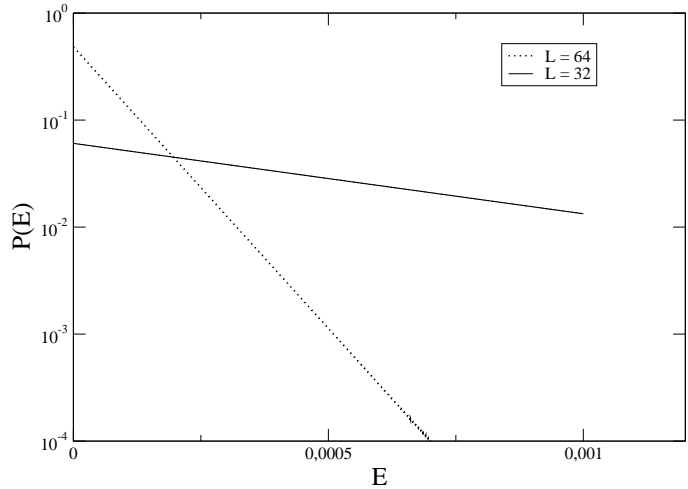}
\end{center}

\caption{Energy distributions for two fixed system sizes, computed via the
assumption that one can equate $L=2^X$ where $X$ is the number of independent
lengthscales (corresponding to the 3d case).}
\label{figa1}
\end{figure}

\begin{figure}
\begin{center}
\includegraphics[width=9cm]{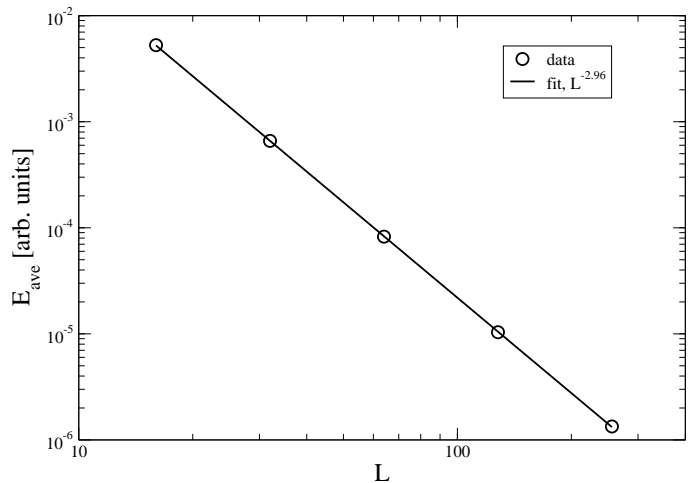}
\end{center}
\caption{Average minimum droplet energy from the droplet
argument, computed via integrating numerically Eq.\ (\ref{eq:PEfinal}) 
and the resulting average for the 3d case. The data implies an inverse
volume-like scaling $\overline{E} \sim 1/V = 1/L^3$.}
\label{figa2}
\end{figure}

The resulting definite
integral, using $l=2^x=e^{x\ln 2}$ ($x=0,\ldots,\ln L/\ln 2$),
 is of the type $\int \exp{(-ax)} \ln (1-b\exp{(-cx)}) dx$
and does not appear to have a simple closed form solution. 
Integrating, with fixed $L$ over $l=1\dots\ln L/\ln 2$ produces
for the probability Figure \ref{figa1}. We see that  in analogy
to the numerical findings for paramagnetic 3d systems the 
minimum excitation energy distribution is exponential. Here,
$a$ in $P_l$ is just chosen at random to a small constant.
In reality, when the droplet argument is valid it should depend
on dimensionality and $h$. Figure \ref{figa2} shows for a range
of $L$ values the resulting average droplet energy $\overline{E}$.
In agreement with the most clear-cut numerical data in 3d, the
scaling is volumelike ($\overline{E} \sim 1/L^3$).

\section{Summary and discussion} 

In this paper, we have first presented an algorithm that calculates first excited states in the random-field Ising model. We showed that a former algorithm developed by Frontera and Vives does not find the true 
first excited states in all cases but and therefore needs to be extended.
When comparing the correct algorithm with the Frontera-Vies algorithm,
 we find that the difference in quantities such as the excitation energy is measurable albeit not drastic for 
  larger systems.

The excitations consist of connected clusters of spins which are reversed with respect
to the ground state.
In our simulations we study mainly the three-dimensional RFIM, but we also present 
some results for two dimensions for comparison.  Interestingly,
 there is an intersection of
the  $\overline{E}(h,L)L^3$ curves for different system sizes 
at a point very close to the phase transition point $h_c$. 
This behavior cannot be found in the two-dimensional case and seems to 
reflect the existence of a phase transition.

The volume distribution $P(h)$ exhibits a peak close to $h_c$ that becomes sharper with growing system size. However, the size distribution $P(V)$ at a fixed $h$ near the peak is relatively broad so that statistical 
errors are large. The fractal dimension of the clusters was determined to be $d_s=2.42(3)$.

Using analytical arguments,
we were able to understand
the behavior of the average excitation energy 
$\overline{E}(h)$ in the extreme cases $h\gg_c$ and $h\ll h_c$.
To further investigate the origins of the observed energies and
their scaling, we have used a droplet argument to derive the 
energy distribution of the minimal excitations. This line of
reasoning is similar to a few others found in the literature
(see again e.g.\ \cite{Middleton01b}, but we are not aware of
any where the whole distribution would have been written down.
The result reproduces well some features of the numerics,
as a volume-like scaling of $\overline{E}$ and the exponential
character of the $P(E)$.
Hence, the such calculated behavior at $h_c$ follows 
our numerical observations.
Additional corrections to scaling close to $h_c$ seem to be
responsible for the observed behavior of the crossing of the 
$\overline{E}(h,L)L^3$ curves.

For future work, it would be interesting to look at other types of low excitations with ground states that do not necessarily produce the FES but still a low energy state. States of this kind can be generated much more efficiently, leading to larger systems that can be simulated. Another possibility would be to analyze excitations in higher dimensional systems (see \cite{Hartmann00} \cite{Middleton02} for results in $d=4$). However, different approaches from the FES algorithm would have to be used for that since we were already restricted to relatively small system sizes in three dimensions.

\section{Acknowledgments}
The authors have received financial support from the {\em
  VolkswagenStiftung} (Germany) within the program ``Nachwuchsgruppen
an Universit\"aten'', and from the European Community via the
DYGLAGEMEM program.
 MJA would like to acknowledge the
support of the Center of Excellence program of the Academy of
Finland.







\addcontentsline{toc}{chapter}{\numberline{}Bibliography}
\bibliographystyle{unsrt}

\bibliography{literatur}

\end{document}